\documentclass[onecolumn,noshowpacs,preprintnumbers]{revtex4}

\usepackage{graphicx,amsmath,array,tabularx,natbib}

\begin{document}
\title{Direct MD simulation of liquid-solid phase equilibria for three-component plasma}
\date{\today}
\newcolumntype{W}{>{\raggedright\arraybackslash}X}
\newcolumntype{Y}{>{\raggedleft\arraybackslash}X}
\newcolumntype{Z}{>{\centering\arraybackslash}X}
\author{J. Hughto}
\email{jhughto@astro.indiana.edu}
\author{C. J. Horowitz}
\email{horowit@indiana.edu}
\author{A. S. Schneider}
\affiliation{Department of Physics and Nuclear Theory Center, Indiana University, Blooomington, IN 47405}

\author{Zach Medin}\email{zmedin@lanl.gov}
\affiliation{Los Alamos National Laboratory, Los Alamos, NM 87545, USA}
\author{Andrew Cumming}\email{cumming@physics.mcgill.ca}
\affiliation{Department of Physics, McGill University, 3600 rue University,
Montreal, QC H3A 2T8, Canada}

\author{D. K. Berry}
\affiliation{University Information Technology Services, Indiana University, Bloomington, IN 47408}

\begin{abstract}
The neutron rich isotope $^{22}$Ne may be a significant impurity in carbon and oxygen white dwarfs and could impact how the stars freeze.  We perform molecular dynamics simulations to determine the influence of $^{22}$Ne in carbon-oxygen-neon systems on liquid-solid phase equilibria.  Both liquid and solid phases are present simultaneously in our simulation volumes.  We identify liquid, solid, and interface regions in our simulations using a bond angle metric.  In general we find good agreement for the composition of liquid and solid phases between our MD simulations and the semi analytic model of Medin and Cumming.  The trace presence of a third component, neon, does not appear to strongly impact the chemical separation found previously for two component carbon and oxygen systems.  This suggests that small amounts of $^{22}$Ne may not qualitatively change how the material in white dwarf stars freezes.  However, we do find systematically lower melting temperatures (higher $\Gamma$) in our MD simulations compared to the semi analytic model.  This difference seems to grow with impurity parameter $Q_{imp}$ and suggests a problem with simple corrections to the linear mixing rule for the free energy of multicomponent solid mixtures that is used in the semi analytic model.  

\end{abstract}

\maketitle

\section{Introduction}
The internal composition of White Dwarf (WD) stars has an impact on their evolution.  Sedimentation of $^{22}$Ne and the release of latent heat of fusion has been shown to delay cooling of WD stars by a few Gyr \cite{garciaberro2010}.  Such modest energy sources can have a large effect because the energy input from nuclear reactions is small.  Therefore, the chemical energy input becomes relevant.  Understanding these additional energy sources can allow for more accurate age determinations of stellar clusters.

The interior of a WD is a Coulomb plasma of ions and a degenerate gas of electrons.  As the star cools this plasma crystallizes.  This crystallization has been observed in recent observations of both globular \cite{winget2009} and open star clusters \cite{garciaberro2011}. The melting temperature observed in the \citeauthor{winget2009} results may also constrain the composition of the WD interior\cite{horowitz2010}.  \citeauthor{horowitz2010} only considered the contribution of carbon and oxygen.

Much of the carbon, nitrogen, and oxygen, originally present in the star, is converted by nuclear reactions to the neutron-rich isotope $^{22}$Ne.  The larger mass to charge ratio of $^{22}$Ne compared to $^{12}$C and $^{16}$O results in a release of gravitational energy as it sinks in the strong gravitational field of the star.  This provides an additional source of energy during WD cooling \cite{bildsten2001,deloye2002}.  Sedimentation of $^{22}$Ne can also affect the $^{56}$Ni yield in type Ia SNe since the excess $^{22}$Ne in the core can modify the electron fraction \cite{timmes2003}.

Liquid-solid phase diagrams for multicomponent plasmas have been determined using Monte Carlo and density-functional techniques.  Binary mixtures have been considered by many groups \cite{segretain1993,ogata1993,ichimaru1988,dewitt2003,dewitt1996}, while three or more components have not been as extensively studied \cite{segretain1996,potekhin2009,potekhin2009b,medin2010,stringfellow1990}.  Often these works determine liquid-solid phase equilibria by comparing liquid and solid free energies that have been calculated separately.  This approach may be sensitive to any small errors in the free energy difference between liquid and solid phases, while also providing no information on the dynamics of the phase transition.

Recently, we have performed direct two-phase molecular dynamics simulations of liquid-solid phase equilibria for carbon-oxygen mixtures in WD stars \cite{horowitz2010,schneider2011}, oxygen-selenium mixtures \cite{schneider2011}, and for a complex 17 component mixture modeling the crust of an accreting neutron star \cite{horowitz2007}.  These simulations have both liquid and solid phases present simultaneously.  This allows a direct determination of the melting temperature, and the composition of the liquid and solid phases from a single simulation.  Systems with an arbitrary number of components can be modeled in this way.

One must address potential systematic errors from finite size and non-equilibrium effects, however.  Finite size effects can be important since an ion deep in the bulk liquid or solid may not be far from an interface.  Larger system sizes must be used to address this.  These two-phase simulations must also be run long enough to ensure that the phases come into thermodynamic equilibrium.  This requires impurities to diffuse through the  solid phase.  However, diffusion in the solid phase is relatively fast since the ions have soft $1/r$ interactions, instead of hard cores, so that the ions can move past one another.  We have extensively studied diffusion in Coulomb crystals in a recent paper \cite{hughto2011}.

If the systematic effects of a direct molecular dynamics simulation are addressed, then this method should yield accurate results.  The systematic errors between the direct MD simulations and the free energy calculations are in principle very different so comparing the results directly should provide an important check on both methods.  However, due to the lack of published results for free energy calculations of multicomponent systems, the only quantitative method currently available for comparing to these calculations is an extrapolation of their results for two-component systems \cite{medin2010}.

In this paper, we perform MD simulations of C/O/Ne mixtures with 27648 ions for three different C/O ratios as well as three different Ne concentrations for a total of nine systems.  This choice allows us to see the neon dependence of the phase diagram across a range of C/O ratios.  We also compare our MD results to an analytic model extrapolating the results of two-component, free energy calculations.  We discuss our MD formalism and analytic model in Section II, present results in Section III, and conclude in Section IV.
\section{Formalism}

We describe our molecular dynamics method in Section IIa, the algorithm to determine liquid vs solid vs interface in Section IIb, and the theoretical model in Section IIc.

\subsection{MD formalism}
The method used in these simulations is similar to our previous liquid-solid equilibria determinations \cite{horowitz2010,schneider2011}.  We consider a three component mixture of $^{12}$C, $^{16}$O, and $^{22}$Ne.  The ions interact via screened Yukawa interactions
\begin{equation}
v_{ij}(r)=\frac{Z_iZ_je^2}{r}e^{-r/\lambda},
\end{equation}
where $Z_i$ and $Z_j$ are the respective charges of the two ions being considered, $r$ is the separation between the ions, and $\lambda$ is the Fermi screening length, which for cold relativistic ions is
\begin{equation}
\lambda^{-1}=2k_F\sqrt{\frac{\alpha}{\pi}}.
\end{equation}
Here, the electron Fermi momentum $k_F$ is $k_F=\left(3\pi^2 n_e\right)^{1/3}$ and $\alpha$ is the fine structure constant.  The electron density $n_e$ is equal to the ion charge density, $n_e=\langle Z\rangle n$, where $n$ is the ion density and $\langle Z\rangle$ is the average (by number of ions).  Our simulations are classical and we have neglected the electron mass (extreme relativistic limit).  Note that electrons can be non relativistic at lower densities and this will modify $\lambda$.  However, our results are not very sensitive to the exact value of $\lambda$, see for example ref. \cite{hamaguchi1996}.

One component plasma simulations can be characterized by the Coulomb parameter $\Gamma=Z^2e^2/aT.$   We characterize our multicomponent system using an average Coulomb parameter, 
\begin{equation}
\label{eq:gamma}
\Gamma=\langle Z^{5/3}\rangle\Gamma_e, 
\end{equation}
where $\Gamma_e=e^2/a_eT$ with the electron sphere radius $a_e=(3/4\pi n_e)^{1/3}$.  For our mixtures of carbon, oxygen, and neon, the values of $a_e$ and $\lambda$ are such that the dimensionless ratio $\kappa\equiv a_e/\left<Z\right>^{1/3}\lambda\lesssim0.4$.  Therefore, we expect the ground state to be a body-centered cubic (bcc) rather than a face-centered cubic (fcc) crystal. \citet{hamaguchi1996} finds $\kappa\geq1.066$, for a one-component plasma, in order for the system to be an fcc crystal.

Time can be measured in our system in units of one over the plasma frequency $\bar{\omega}_p$.  For a one component plasma, the plasma frequency is
\begin{equation}
\omega_p=\left[\frac{4\pi Z^2 e^2 n}{M}\right]^{1/2},
\end{equation}
where $M$ is the mass of the ion.  For a mixture, we choose to define an average plasma frequency $\bar{\omega}_p=\left(4\pi \langle Z\rangle^2 e^2 n/\langle M\rangle\right)^{1/2}$.

\subsection{Interface finding algorithm}
\label{sec:bond}
Determining whether a cluster of ions is a liquid or a bcc solid is simple when visually inspected, however this determination is difficult to obtain numerically.  For an entire system, phase determination can be accomplished by computing the global order parameter $Q_6$\cite{steinhardt83}. In this work, we use the prescription laid out by \citet{tenwolde96} to determine whether individual ions are liquid-like or solid-like.

For each ion $i$, an ion $j$ is defined as a neighbor if it is within a given radius $r_{min}$, as defined by the first minimum in the pairwise correlation function $g(r)$.  The vectors $\hat{\textbf{r}}_{ij}$ joining neighbors are called bonds.  The direction of these vectors can be described by $\theta_{ij}$ and $\phi_{ij}$ in the frame of ion $i$.  The local structure around ion $i$ can be characterized by a spherical vector $\bar{q}_{lm}(i)$, 
\begin{equation}
\bar{q}_{lm}(i)=\frac{1}{N_b(i)}\sum_{j=1}^{N_b(i)}Y_{lm}(\theta_{ij},\phi_{ij}),
\end{equation}
where $N_b(i)$ is the number of ions bonded with ion $i$ and $Y_{lm}(\theta_{ij},\phi_{ij})$ is a spherical harmonic.

These local order parameters are large in both the solid and the liquid.  The global order parameter $Q_6$ is calculated from an average over all of the $N$ ions,
\begin{eqnarray}
q_{6m}&=&\frac{1}{N}\sum_{i=1}^N\bar{q}_{6m}(i),\\
Q_6&=&\left[\frac{4\pi}{13}\sum_{m=-6}^6\left|q_{6m}\right|^2\right]^{1/2}.
\end{eqnarray}
This is large in the solid due to the fact that the $\bar{q}_{6m}(i)$ add up coherently.  In the liquid, $\bar{q}_{6m}(i)$ add incoherently, so $Q_6$ is near zero.  This coherence is exploited to determine local order.  For each $\bar{q}_{6m}(i)$ a normalization is applied,
\begin{equation}
\tilde{q}_{6m}(i)\equiv\frac{\bar{q}_{6m}(i)}{\displaystyle\left[\sum_{m=-6}^6\left|\bar{q}_{6m}(i)\right|^2\right]^{1/2}}
\end{equation}
A dot product can now be defined of the vectors $\textbf{q}_6$ for neighboring particles $i$ and $j$,
\begin{equation}
\textbf{q}_6(i)\cdot\textbf{q}_6(j)\equiv\sum_{m=-6}^6\tilde{q}_{6m}(i)\tilde{q}_{6m}^*(j).
\end{equation}
By construction, $\textbf{q}_6(i)\cdot\textbf{q}_6(i)=1$.

We use the same criterion as \citet{tenwolde96} for determining whether two particles are connected, namely $\textbf{q}_6(i)\cdot\textbf{q}_6(j)>0.5$.  This criterion will be met for most of the bonds in the solid.  In the liquid, two neighbors may be in phase and considered connected, but that is certainly not true for all of the neighbors.  Therefore, we use a threshold on the number of connections to determine if an ion is solid-like or liquid-like.  If an ion has more than seven connections, then it is tagged as solid-like. If it has seven or fewer connections it is liquid-like.  For a perfect bcc crystal, the number of connections per ion is 14.

Now that each ion is tagged as either solid-like or liquid-like, the interface in our two-phase simulations can be found.  Deep in the solid, a vast majority of the ions within a certain radius of a given ion are identified as solid-like.  In the bulk of the liquid a similar majority of ions are identified as liquid-like.  Along the interface, there is a mixture of solid-like and liquid-like ions.  For this reason, we tag an ion as being in the solid or liquid if a large majority ($>0.95$) of ions within two lattice spacings are in the same phase.  If this criterion is not met, then the ion is determined to be in the interface.

\begin{figure}[ht]
\centering
\includegraphics[width=3.5in]{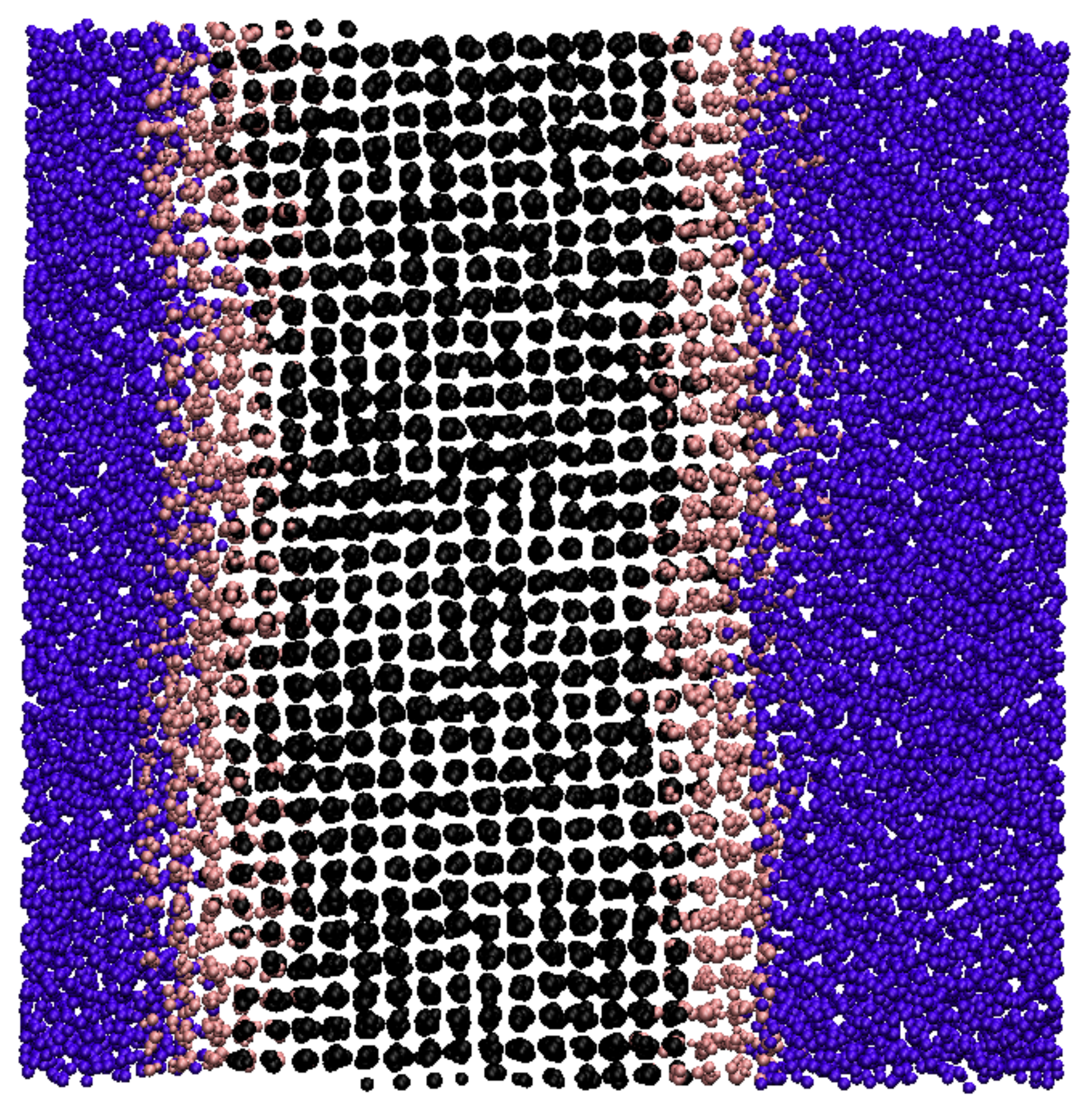}
\caption{\label{fig:vmd}(Color on line)Automated interface finding algorithm results.  Ions identified as being in the solid, liquid, or interface are colored black, purple, or pink, respectively.}
\end{figure}

Requiring a smaller majority, say $>75\%$, to define liquid and solid regions may lead to a thinner interface region.  This could slightly increase finite size effects for the composition of the solid and liquid because we expect the interface region to have a composition, in general, intermediate between that of the solid and liquid phases.  Therefore we choose an interface definition to yield a reasonably thick interface as shown in Fig. \ref{fig:vmd}.  Ions determined to be in the interface are found where one would expect them, along the border of the solid and liquid phases.  Note that many of the ions determined to be part of the interface look to be part of the solid by eye.

\subsection{Semi analytic extrapolation of results from previous works}
\label{sec:theory}

There are a few calculations of three-component plasmas and in
particular the carbon-oxygen-neon system \citep{ogata1993,segretain1996}, but
it is difficult to make quantitative comparisons with these works. On
the other hand, there are several quantitative results for one- and
two-component plasmas \citep[e.g.][]{ogata1993, dewitt2003}. Using the method
described in \citet{medin2010}, hereafter MC10, we can extrapolate
the one- and two-component results for comparison with the C/O/Ne
system described in this paper. A full description of the
extrapolation method is given in MC10; here we outline the basic
algorithm.

A multicomponent plasma (MCP) can be characterized by the charge $Z_i$
and fraction composition $x_i = N_i/N$ of each ion species and the average
Coulomb coupling parameter, $\Gamma$ [Eq. (\ref{eq:gamma})].  For a given $\vec{Z} = \{Z_i\}$, $\vec{x} = \{x_i\}$, and
$\Gamma$, the equilibrium state of the mixture is fixed. This state,
the state of lowest free energy, may be a pure liquid phase, a pure
solid phase, or some combination of liquid and solid phases. For a
given $\Gamma$, an MCP of composition $\vec{x}$ lies in a transition
state between two phases of compositions $\vec{a}$ and $\vec{b}$ if it
can be formed from a linear combination of the two phases and its free
energy as a pure phase is greater than the total free energy of the
combined phases; i.e., if
\begin{equation}
A\vec{a}+(1-A)\vec{b} = \vec{x}
\label{eq:x}
\end{equation}
and
\begin{equation}
Af_a+(1-A)f_b < f_x
\end{equation}
for some $0 < A < 1$, where $f_x$ is the free energy per particle of
composition $x$, etc. The extrapolation method proceeds in two parts:
First, the free energies of both the liquid and the solid phases of
the MCP are calculated from the relevant one- and two-component
values. Second, minimization is performed over (liquid and solid)
composition space to find the mixture $\vec{a},\vec{b},A$ that gives
the lowest total free energy for that composition, or to show that
pure $\vec{x}$ is lower in energy than any other mixture with the same
average composition.

The free energy of the liquid phase of a multicomponent plasma is
well described by the linear mixing rule (but see Appendix B of MC10);
i.e.,
\begin{align}
f_l^{\rm MCP}(\Gamma_1,x_1, & \ldots,x_{m-1}) \nonumber\\
\simeq & \sum_{i=1}^m x_i \left[f_l^{\rm OCP}(\Gamma_i) + \ln\left(x_i \frac{Z_i}{\langle Z \rangle}\right)\right]
\label{eq:flMCP}
\end{align}
where $\langle Z \rangle = \sum_{i=1}^m x_iZ_i$ is the average ion charge 
and $m$ is the number of ion species. The free energy of the solid phase of the MCP is
\begin{align}
f_s^{\rm MCP}(\Gamma_1,x_1, & \ldots,x_{m-1}) \nonumber\\
 \simeq & \sum_{i=1}^m x_i \left[f_s^{\rm OCP}(\Gamma_i) + \ln\left(x_i \frac{Z_i}{\langle Z \rangle}\right)\right] \nonumber\\
 & + \Delta f_s(\Gamma_1,x_1,\ldots,x_{m-1}) \,.
\label{eq:fsMCP}
\end{align}
The expression we use for the deviation of the solid from linear
mixing $\Delta f_s$ is that given in Ref. \cite{ogata1993}; another
expression for $\Delta f_s$ can be found in Ref. \cite{dewitt2003} \citep[see also][]{segretain1993,segretain1996}. Here, $f_l^{\rm OCP}(\Gamma_i)$ and $f_s^{\rm
OCP}(\Gamma_i)$ are the free energies of a one-component plasma \citep[as taken from, e.g.][]{stringfellow1990,dewitt2003}.

For a multi-component plasma, assuming $\vec{a}$ is in the liquid
state and $\vec{b}$ is in the solid state the minimization equations
to solve are
\begin{align}
f_l(\vec{a}) + \frac{df_l}{dx_i}(\vec{a}) - \vec{a} \cdot \nabla f_l(\vec{a}) & = f_s(\vec{b}) + \frac{df_s}{dx_i}(\vec{b}) - \vec{b} \cdot \nabla f_s(\vec{b}) \,, \nonumber\\
i & \in [1,m-1]
\label{eq:MCP1m1}
\end{align}
and
\begin{equation}
f_l(\vec{a}) - \vec{a} \cdot \nabla f_l(\vec{a}) = f_s(\vec{b}) - \vec{b} \cdot \nabla f_s(\vec{b})
\label{eq:MCPm}
\end{equation}
(see equations 28 and 29 of MC10). With Eqs.~(\ref{eq:MCP1m1}) and
(\ref{eq:MCPm}) we must make a choice: Do we solve for $\Gamma$
given the fraction $0<A<1$ of the mixture in state $\vec{a}$, or do we
solve for $A$ given $\Gamma$?  As will be discussed below, we choose
the former for this paper. If we are given an average composition
$\vec{x}$ and the fraction $0<A<1$ of the solution in state $\vec{a}$
(or the fraction $1-A$ in state $\vec{b}$), we can solve for
$\Gamma$ and the compositions of both the liquid and solid mixtures
in equilibrium. We have $2m-1$ unknowns, $a_1,\ldots,a_{m-1}$,
$b_1,\ldots,b_{m-1}$, and $\Gamma$; but in addition to the $m$
equations Eqs.~(\ref{eq:MCP1m1}) and (\ref{eq:MCPm}) above we have the
$m-1$ equations
\begin{equation}
Aa_i+(1-A)b_i = x_i \,, \qquad i \in [1,m-1] \,.
\label{eq:Aeq}
\end{equation}
(note that $\sum a_i=1$, $\sum b_i=1$).

\begin{table*}
\caption{Results using the theoretical model in Section \ref{sec:theory}. This
approach finds the lowest energy liquid-solid mixture where $x_i =
Ax_i^l + (1-A)x_i^s$. The total composition $\{x_i\}$ and the liquid
fraction of the system $A$ are chosen to match values from our
simulation runs.}
\centering
\begin{tabularx}{0.75\textwidth}{c@{\quad} Z Z | Z | Z | Z Z Z | Z Z@{\quad} c}
\multicolumn{4}{c|}{Total} & & \multicolumn{3}{c}{Solid} & \multicolumn{3}{c}{Liquid} \\
$x_{\rm C}$ & $x_{\rm O}$ & $x_{\rm Ne}$ & $\Gamma$ & $A$ & $x_{\rm C}^s$ & $x_{\rm O}^s$ & $x_{\rm Ne}^s$ & $x_{\rm C}^l$ & $x_{\rm O}^l$ & $x_{\rm Ne}^l$ \\
\hline
0.20 & 0.60 & 0.20 & 226.79 & 0.61 & 0.112 & 0.668 & 0.220 & 0.258 & 0.555 & 0.187 \\
0.22 & 0.68 & 0.10 & 217.38 & 0.57 & 0.141 & 0.754 & 0.105 & 0.280 & 0.624 & 0.096 \\
0.24 & 0.74 & 0.02 & 209.12 & 0.57 & 0.154 & 0.826 & 0.020 & 0.304 & 0.676 & 0.020 \\
0.40 & 0.40 & 0.20 & 252.61 & 0.56 & 0.316 & 0.468 & 0.216 & 0.465 & 0.347 & 0.188 \\
0.45 & 0.45 & 0.10 & 239.31 & 0.62 & 0.358 & 0.530 & 0.112 & 0.507 & 0.400 & 0.093 \\
0.49 & 0.49 & 0.02 & 227.78 & 0.57 & 0.421 & 0.561 & 0.018 & 0.541 & 0.437 & 0.022 \\
0.60 & 0.20 & 0.20 & 265.44 & 0.52 & 0.553 & 0.235 & 0.212 & 0.644 & 0.167 & 0.189 \\
0.68 & 0.22 & 0.10 & 245.54 & 0.50 & 0.648 & 0.250 & 0.102 & 0.712 & 0.190 & 0.098 \\
0.68 & 0.22 & 0.10 & 243.91 & 0.72 & 0.629 & 0.266 & 0.105 & 0.700 & 0.202 & 0.098 \\
0.74 & 0.24 & 0.02 & 219.13 & 0.61 & 0.731 & 0.260 & 0.009 & 0.746 & 0.227 & 0.027 \\
\end{tabularx}
\label{tab:extrap}
\end{table*}

Using the above method we generate results for the C/O/Ne
system. These results, presented in Table~\ref{tab:extrap}, represent
our best estimate of what Ogata et al. and other groups with similar
free energy calculations would have found had they run their
calculations for C/O/Ne, and in that sense allows us to compare our
simulation results to those of previous works. For each result, the
total composition $\{x_i\}$ and the liquid fraction of the system $A$
are chosen to match values from a particular simulation run; given
these values we calculate $\Gamma$ and the liquid and solid
compositions, $x_{\rm C}^l,x_{\rm O}^l,x_{\rm Ne}^l$ and $x_{\rm
C}^s,x_{\rm O}^s,x_{\rm Ne}^s$, respectively. Ideally, the value of
$A$ for each comparison is that of the liquid fraction used in the
simulation run. However, the finite size of the simulation provides
some ambiguity to the problem. In the simulations there is an
interface region which is neither solid nor liquid that affects the
total composition; in the semianalytic method discussed here the
interface is assumed to be negligibly small (the system is assumed to
be very large). To get around this ambiguity we effectively ignore the
interface region and choose $A$ to solve
\begin{equation}
Ax_{\rm C}^l+(1-A)x_{\rm C}^s = x_{\rm C} \,.
\end{equation}
Note that a different value for $A$ would be obtained had we chosen to
solve for it in terms of $x_{\rm O}$ or $x_{\rm Ne}$. However, $A$
calcluated in terms of $x_{\rm O}$ is very similar to that of $x_{\rm
C}$ and leads to differences in results of no more than 5\%; $A$
calculated in term of $x_{\rm Ne}$ is also similar except when $x_{\rm
Ne} \ll x_{\rm C,O}$.

Figure~2 of Ref. \cite{horowitz2010} helps show why we used the $A$ values, and
not the $\Gamma$ values, from our simulations to make comparisons
with previous works. Because of systematic differences between our
direct MD simulations and the semianalytic extrapolation, for certain
initial compositions and values of $\Gamma$ (or equivalently,
$T/T_C$ in Horowitz 2010) the final liquid-solid ratios from these two
calculations are qualitatively different. For example, there are
regions of the phase diagrams at large $\Gamma$ where mixtures are
completely solid according to the extrapolation results but are
liquid-solid according to the simulation results. By fixing the liquid
fraction $A$ instead of $\Gamma$ we are guaranteed to find a
liquid-solid mixture using the extrapolation method, though it will
not necessarily be at the same $\Gamma$ as the simulation result.

\section{Results}
We now present results for liquid-solid equilibria for multiple mixtures of $^{12}$C, $^{16}$O, and $^{22}$Ne.  We intend to explore the effect of various trace neon concentrations $x_{Ne}=0.02,\;0.10,\;0.20$ in different ratios of carbon to oxygen, $3:1,\;1:1,\;1:3$.  For these compositions, the ratio of carbon to oxygen has a larger qualitative effect than different neon concentrations, so our runs will be grouped based on the carbon to oxygen ratio.

\subsection{Runs with 1 to 3 carbon to oxygen ratio}
\label{sec:13}

Our first set of simulations had a carbon to oxygen ratio of 1/3 and we used $x_{Ne}=0.02,$ $0.10,$ and $0.20$. We randomly place 3456 ions in a simulation volume at a somewhat high temperature $\Gamma\sim150$ relative to the expected melting temperature $175<\Gamma_c\lesssim300$.  We then make four copies of this system and this constitutes the initial liquid.  We then take another 3456 ions at a lower temperature $\Gamma\sim300$ and allow the system to crystallize.  After evolving the solid system to allow it to roughly equilibrate, $t\sim60 000/\bar{\omega}_p$, we make four copies of the system to make the total solid initial configuration.  The liquid and solid layers are then joined to make a total system size of 27648 ions.

These initial configurations are not equilibrated for a number of reasons.  First, since we combined a total of eight sets of initial conditions in building each large system, the energies along the boundaries may be high.  There may be ions that are placed close together.  The crystal is also not fully equilibrated with respect to structure and composition.  These systems may take considerable time to fully equilibrate.

We then evolve the system in small time steps, $\Delta t\sim1/9\bar{\omega}_p$.  The temperature is adjusted to maintain roughly 50\% liquid and 50\% solid.  In Section IIb we showed that the phase fraction is a difficult determination by eye when including the interface and generally have 50\% liquid, 30\% solid, and 20\% interface ions.  The system is then evolved for a total time of at least $t=2.5\times10^6/\bar{\omega}_p$ to allow the system to fully equilibrate and allow the impurities to diffuse through the system.  These simulations were performed on the Cray XT5 system Kraken at the National Institute for Computational Sciences.

\begin{figure}[ht]
\centering
\includegraphics[width=6.5in]{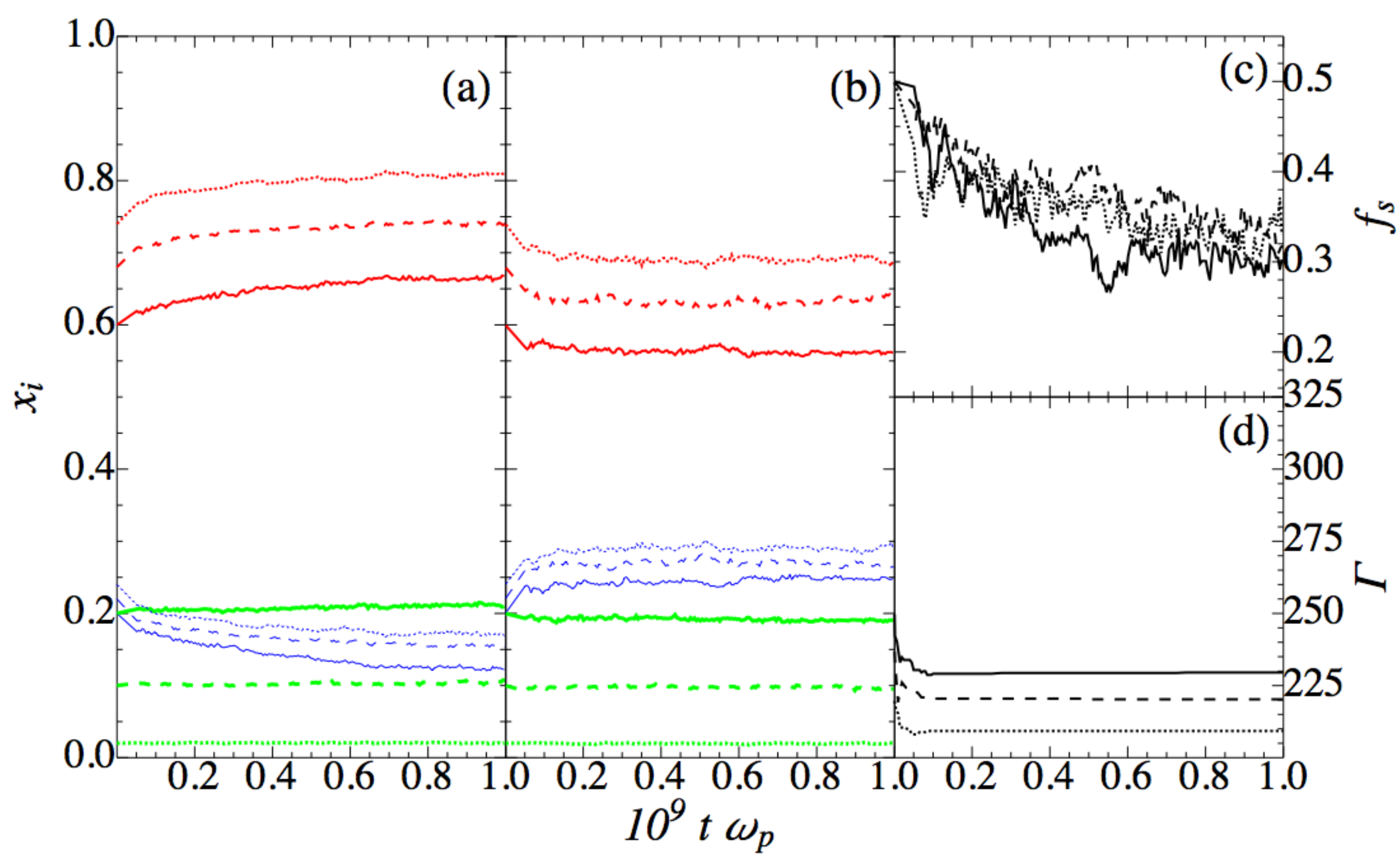}
\caption{\label{fig:CONe_1_3}(Color on line) (a) Composition of the solid region of an MD simulation vs simulation time $t\,\omega_p$.  The blue, red, and green lines indicates carbon, oxygen, and neon number fractions, respectively.  The solid, dashed, and dotted lines correspond to simulations with $x_C/x_O=1/3$ and $x_{Ne}=0.20,$ $0.10,$ and $0.02$, respectively. (b) Composition of the liquid region of an MD simulation vs simulation time.  These data are presented in the same fashion as the left-hand panel.  (c) Fraction of the system that is solid $f_s$ vs $t\,\omega_p$ for $x_{Ne}=0.20$ (solid), $0.10$ (dashed), and $0.02$ (dotted).  All of our simulations have roughly 20\% of the simulation volume determined to be in an interface region.  In order to compute the liquid fraction, on must subtract the solid fraction plotted from $\sim0.8$. (d) $\Gamma$ vs $t\,\omega_p$ for $x_{Ne}=0.20$ (solid), $0.10$ (dashed), and $0.02$ (dotted).  All of these simulations are done with 27648 ions.}
\end{figure}

Figure \ref{fig:CONe_1_3} shows the results of the first set of runs with $x_C/x_O=1/3$, and $x_{Ne}=0.20,$ $0.10,$ and $0.02$.  The number fraction of each ion species in the solid as a function of time is plotted in the left panel.  The center panel contains the number fraction of the ion species in the liquid plotted as a function of time.  The top right panel plots the fraction of the system that is solid.  The interface region for all of our systems is roughly constant and about 20\% of the volume.  In order to determine liquid fraction $f_l$, one must subtract $f_s$ from $\sim0.8$.  We have also plotted $\Gamma$ as a function of time in the bottom right panel.  The scales remain constant across Figures \ref{fig:CONe_1_3}, \ref{fig:CONe_1_1}, and \ref{fig:CONe_3_1} in order to directly compare runs with different compositions.

As the run starts, $\Gamma$ was kept high (low temperature) in order to keep the badly non-equilibrated solid frozen.  As the system rapidly equilibrates, the temperature was then raised towards its final equilibrated value.  The temperature then fluctuated at about the half percent level for the rest of the run.

We expect the solid to be enriched in oxygen and the liquid to be enriched in carbon based on our recent carbon-oxygen phase diagram work \cite{horowitz2010,schneider2011}.  The results shown in Figure \ref{fig:CONe_1_3} suggest that neon as an impurity does not qualitatively affect these results.

\subsection{Runs with 1 to 1 carbon to oxygen ratio}
\label{sec:11}

\begin{figure}[ht]
\centering
\includegraphics[width=6.5in,angle=0,clip=true]{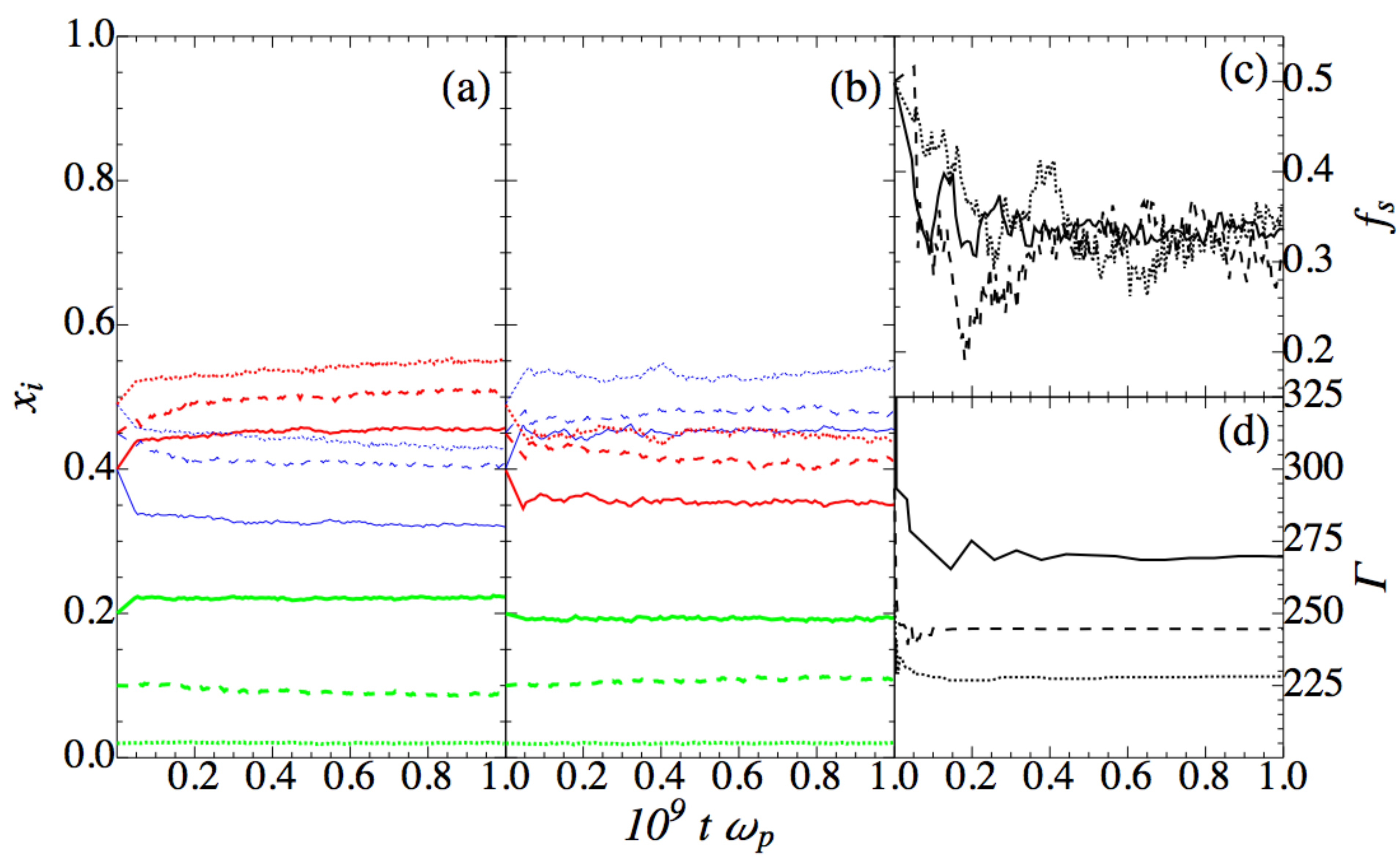}
\caption{\label{fig:CONe_1_1} (Color on line) Compositions and phase fractions as per Fig. \ref{fig:CONe_1_3} except for initial compositions with $x_C/x_O=1/1$.}
\end{figure}

Our second set of runs use a 1/1 carbon to oxygen ratio with $x_{Ne}=0.02,$ $0.10,$ and $0.20$.  We start these systems in a very similar way to the 3/1 carbon oxygen ratio systems.  The history of these three runs is shown in Figure \ref{fig:CONe_1_1}.  We expect, once again, that the solid will be enriched in oxygen and the liquid enriched in carbon.  As in the previous sets of runs, the addition of neon does not qualitatively change these results.  

\subsection{Runs with 3 to 1 carbon to oxygen ratio}
\label{sec:31}

\begin{figure}[ht]
\centering
\includegraphics[width=6.5in,angle=0,clip=true]{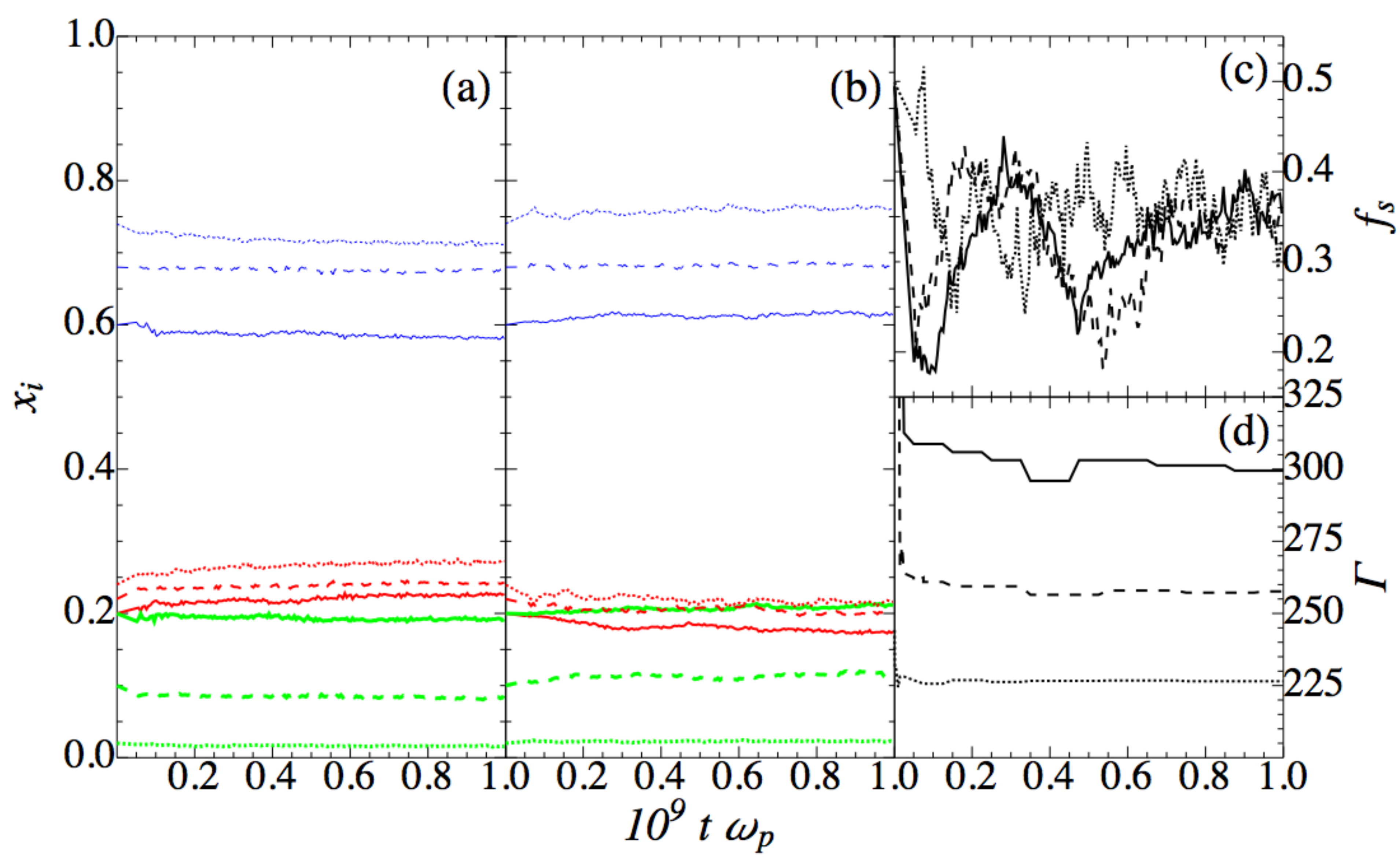}
\caption{\label{fig:CONe_3_1} (Color on line) Compositions and phase fractions as per Fig. \ref{fig:CONe_1_3} except for initial compositions with $x_C/x_O=3/1$.}
\end{figure}

Our final set of runs has an overall carbon to oxygen ratio of 3 to 1.  We started these runs in a very similar way to all other runs described.  The history of these three runs is shown in Figure \ref{fig:CONe_3_1}.

Our previous carbon-oxygen phase diagram suggest that there should be no difference in this ratio between the liquid and solid for carbon and oxygen.  Now neon is enriched in the liquid phase in all cases.

\subsection{Collected results and supplemental run}
\label{sec:phasediagram}

\begin{figure}[ht]
\centering
\includegraphics[width=3.6in,angle=0,clip=true]{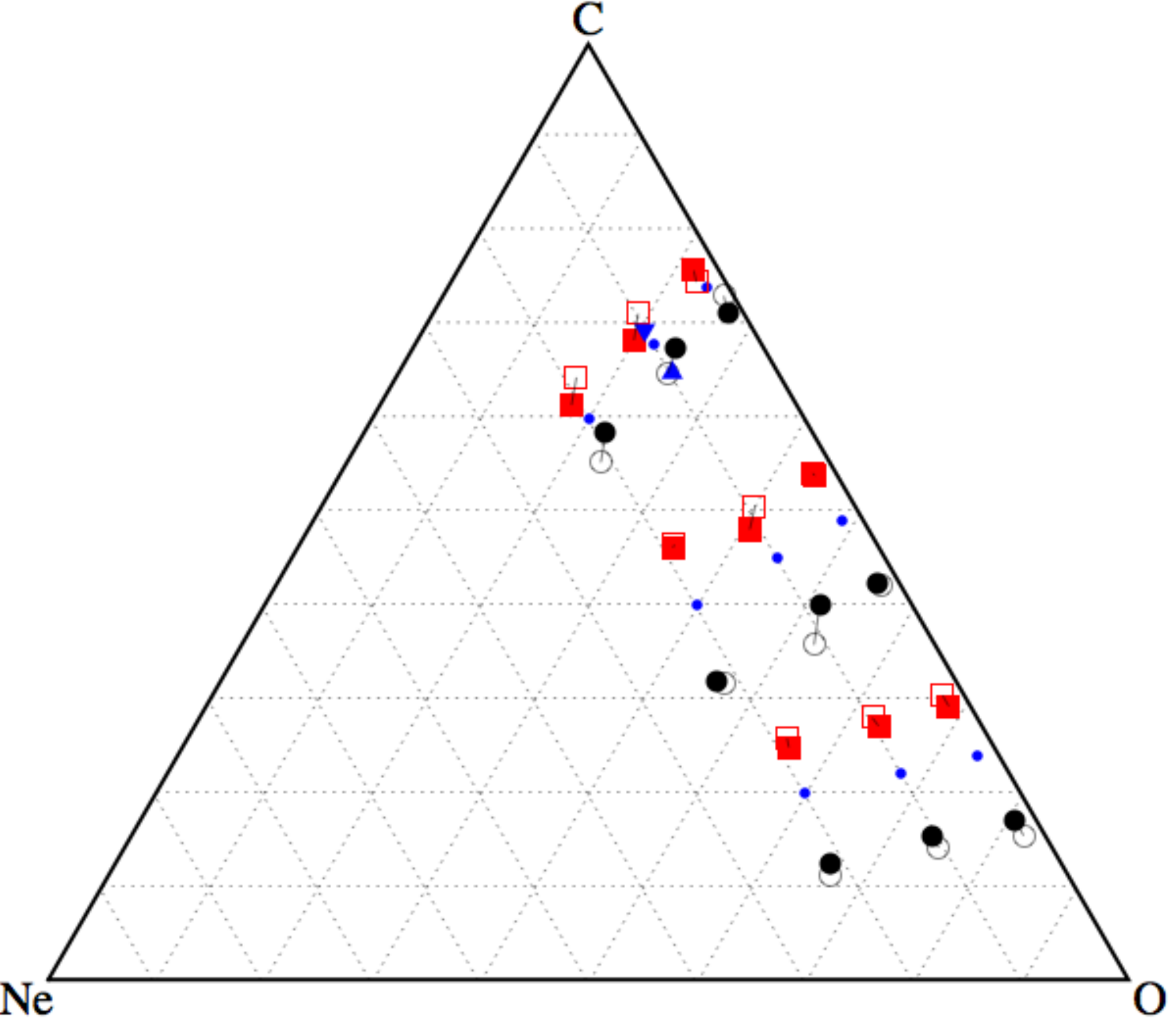}
\caption{\label{fig:phasediagram}(Color on line) Three component liquid-solid equilibria results.  Vertices of triangle plot are 100\% of the labeled ion species.  For example, the vertex labeled $C$ would correspond to a simulation with $x_C=1$ and the line connecting the other two vertices would be $x_C=0$.  Small blue points are initial configurations.  Filled circles(squares) show solid(liquid) equilibrated results.  Open circles(squares) are from theoretical model.  The blue triangles are results of the larger run where the upright(inverted) triangle is the solid(liquid).}
\end{figure}

We now present the collected results from Sections \ref{sec:13}, \ref{sec:11}, \ref{sec:31}.  The final compositions and final temperature of the runs are listed in Table \ref{table:phase}.  These data are shown on a ternary plot in Figure \ref{fig:phasediagram}.  A point on this plot corresponds to the composition $(x_C,x_O,x_{Ne})$ subject to $x_C+x_O+x_{Ne}=1$.  The number fraction of any given species is $x_i=1$ at the labeled vertex and $x_i=0$ on the line opposite it.  The solid symbols are the data from the MD runs and the open symbols are the model predictions.  The fact that the squares and circles lie close to lines of constant neon concentration shows that the neon abundance does not strongly impact the shape of the CO phase diagram.

The 3:1 carbon to oxygen ratio runs have the largest discrepancy between the MD runs and the theoretical model.  In order to determine whether this is an actual difference between what was predicted and what was run or if the difference is due to statistical fluctuations, we performed another run for $x_{C,O,Ne}=0.68,0.22,0.10$ with twice as many ions, $N=55296$, in a rectangular box with $L_z=2L_x=2L_y$, where $L_i$ are the box lengths in the three directions.  We started this system with a biased composition as was done in Ref. \cite{schneider2011}.  Instead of starting with the liquid and solid at $x_{C,O,Ne}=0.68,0.22,0.10$, we started with the composition predicted in the theoretical model of $x_{C,O,Ne}^s=0.648,0.250,0.102$ in the solid and $x_{C,O,Ne}^l=0.712,0.190,0.098$ in the liquid.  The system was then run to $t=2.5\times10^6/\bar{\omega}_p$.  These results are the blue triangles in Figure \ref{fig:phasediagram}.  Notice how even though the initial conditions agreed with the model predictions, this MD simulation still moved towards our previous $N=27648$ ion MD compositions.

\begin{figure}[ht]
\centering
\includegraphics[width=5in,angle=0,clip=true]{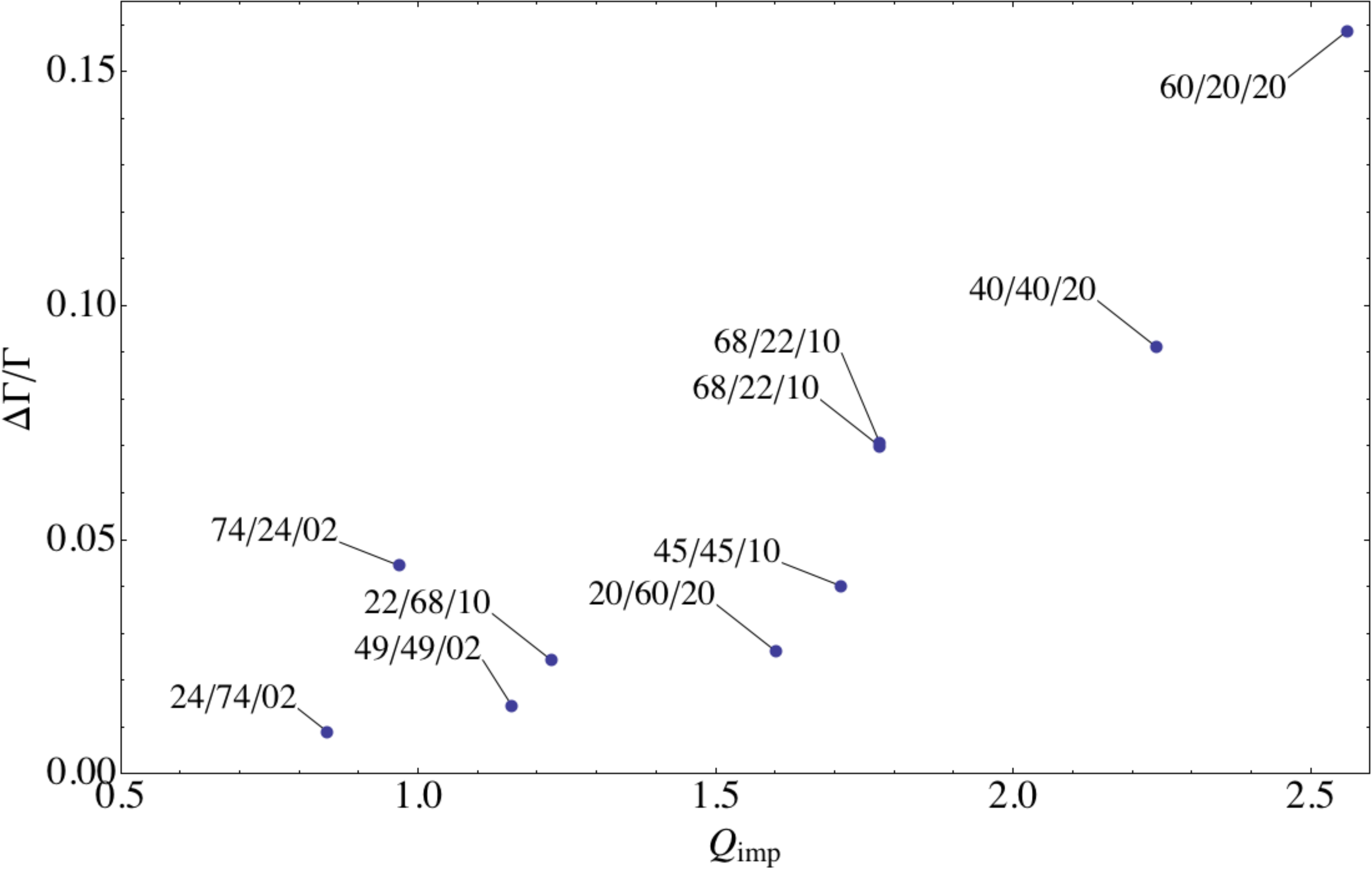}
\caption{\label{fig:Q_imp}(Color on line) Comparison $\Delta \Gamma=\Gamma_{MD}-\Gamma_m$ between the MD simulation predictions for Coulomb parameter $\Gamma_{MD}=\Gamma$ of Section \ref{sec:phasediagram} and the model results $\Gamma_m$ of Section \ref{sec:theory}, versus impurity parameter $Q_{imp}$, see Eq. \eqref{Eq.qimp}.  The carbon, oxygen, and neon number fractions, in \%, of each point are labeled.}
\end{figure}

We now test the agreement on the melting point, $\Gamma$, between our MD simulations and the theoretical model.  Note that the model predictions were determined using our MD data for the liquid fraction, $A$.  In general, the model predicts lower $\Gamma$ values by 5-20\%, see Table \ref{tab:extrap}, than do our MD simulations, see Table \ref{table:phase} and Fig. \ref{fig:Q_imp}.    This difference appears to be correlated with the impurity parameter $Q_{imp}$,
\begin{equation}
Q_{imp}=\sum_{i={\rm C, O, Ne}}x_i(Z_i-\langle Z \rangle)^2\, .
\label{Eq.qimp}
\end{equation}
This parameter measures the dispersion in charge of the ions.  For systems with a small number of impurities (small $x_{Ne}$) there is good agreement between model and simulations.  However the difference is larger for systems with a large number of impurities (in general large $x_{Ne}$).  This is illustrated in Figure \ref{fig:Q_imp}.  This difference suggests a possible problem with the correction to the linear mixing rule for the free energy of the solid phase that is used by the model, see $\Delta f_s$ in Eq. \eqref{eq:fsMCP} that is taken from Ref. \cite{ogata1993}.  For example the assumed bilinear form (see MC10 equation 13) may be too simple.   In future work we will calculate the free energy of solid C, O, Ne mixtures, using single phase MD simulations, and compare to the assumed linear mixing rule corrections.

\begin{table}[ht]
\begin{center}
\caption{\label{table:phase}Composition of the total system, solid, and liquid phases for the nine MD simulations using molecular dynamics.  The Coulomb parameter $\Gamma$ is determined by the final temperature and the total composition of the run, see Eq. \ref{eq:gamma}.}
\begin{tabular}{lll|l| c| lll| lll}
\multicolumn{4}{c}{Total} & & \multicolumn{3}{c}{Solid} & \multicolumn{3}{c}{Liquid}\\
$x_C$&$ x_O$&$ x_{Ne}$&$N_{ion}$&$\Gamma$&$x_C^s$&$x_O^s$&$x_{Ne}^s$&$x_C^l$&$x_O^l$&$x_{Ne}^l$\\
\hline
0.20&0.60&0.20&27648&232.75&0.125&0.661&0.214 &0.249&0.561&0.190 \\
0.22&0.68&0.10&27648&222.69&0.153&0.743&0.105 &0.271&0.633&0.096 \\
0.24&0.74&0.02&27648&211.02&0.170&0.810&0.0196&0.292&0.687&0.0201\\
0.40&0.40&0.20&27648&275.66&0.320&0.460&0.220 &0.462&0.348&0.191 \\
0.45&0.45&0.10&27648&248.93&0.400&0.514&0.086 &0.481&0.409&0.110 \\
0.49&0.49&0.02&27648&231.12&0.424&0.556&0.0202&0.539&0.440&0.0204\\
0.60&0.20&0.20&27648&307.58&0.585&0.224&0.191 &0.614&0.177&0.208 \\
0.68&0.22&0.10&27648&262.95&0.676&0.242&0.082 &0.684&0.200&0.116 \\
0.68&0.22&0.10&55296&261.01&0.652&0.252&0.096 &0.691&0.207&0.102 \\
0.74&0.24&0.02&27648&229.22&0.712&0.273&0.0149&0.758&0.218&0.0233\\
\end{tabular} 
\end{center}
\end{table}

\section{Conclusion}
We have performed molecular dynamics simulations to determine the influence of $^{22}$Ne in carbon-oxygen-neon systems on liquid-solid phase equilibria.  Both liquid and solid phases are present simultaneously in the simulation volume.  We identified liquid, solid, and interface regions in our simulations using a bond angle metric described in Section \ref{sec:bond}.  This is the latest in a series of papers on liquid-solid equilibria both for complex multicomponent rp ash systems on accreting neutron stars \cite{horowitz2007} and for two component carbon-oxygen systems in cooling white dwarfs \cite{horowitz2010,schneider2011}.

In general we find good agreement for the composition of liquid and solid phases between our MD simulations and the semi analytic model of Medin and Cumming \cite{medin2010}, see Fig. \ref{fig:phasediagram}.  The trace presence of a third component, neon, does not appear to strongly impact the chemical separation found previously for two component carbon and oxygen systems.  This suggests that the presence of small amounts of $^{22}$Ne does not qualitatively change how the material in white dwarf stars freezes.  This is consistent with assumptions in recent calculations of white dwarf evolution\cite{althaus2012} but appears to be at odds with Segretain \cite{segretain1996}.  

However, we do find systematically lower melting temperatures by 5-20\% (higher $\Gamma$) in our MD simulations compared to the semi analytic model, see Fig. \ref{fig:Q_imp}.  This difference seems to grow with impurity parameter $Q_{imp}$.  This suggests a problem with the simple corrections to the linear mixing rule for the free energy of multicomponent solid mixtures that is used in the semi analytic model.  These simple corrections may have been fit to simulations involving only a small number of ions and where the distribution of impurities may not have been fully equilibrated.  Alternatively, the model assumes corrections to linear mixing for the three component system can be written as a sum of pairwise terms and this assumption may be inadequate.

To investigate these differences we are performing extensive single phase MD simulations of two and three component solid mixtures.  These simulations involve 27648 or more ions and long simulation times.  We are calculating radial distribution functions and static structure factors.  In addition we are studying the diffusion of impurities and defects in these solid mixtures \cite{hughto2011}.  Not only do impurities occupy substitutional or interstitial positions in the crystal lattice, but the lattice itself can deform locally to accommodate impurities.  This may be because the $1/r$ Coulomb interaction has no intrinsic length scale.  Therefore bond lengths are free to deform locally, depending on the charges of individual ions, as long as an average bond length is reproduced over large distances.  In future work we will report on the structure of solid mixtures and on their free energies.  

This research was supported in part by DOE grants DE-FG0287ER40365 and DE-AC5206NA25396, by the National Science Foundation through XSEDE (eXtreme Science and Engineering Discovery Environment) resources provided by the National Institute for Computational Sciences under grant TG-AST100014, and by the Natural Sciences and Engineering Research Council of Canada and the Canadian Institute for Advanced Research.

\bibliographystyle{apsrev}
\bibliography{CONe}

\begin{thebibliography}{24}
\expandafter\ifx\csname natexlab\endcsname\relax\def\natexlab#1{#1}\fi
\expandafter\ifx\csname bibnamefont\endcsname\relax
  \def\bibnamefont#1{#1}\fi
\expandafter\ifx\csname bibfnamefont\endcsname\relax
  \def\bibfnamefont#1{#1}\fi
\expandafter\ifx\csname citenamefont\endcsname\relax
  \def\citenamefont#1{#1}\fi
\expandafter\ifx\csname url\endcsname\relax
  \def\url#1{\texttt{#1}}\fi
\expandafter\ifx\csname urlprefix\endcsname\relax\def\urlprefix{URL }\fi
\providecommand{\bibinfo}[2]{#2}
\providecommand{\eprint}[2][]{\url{#2}}

\bibitem[{\citenamefont{{Garc{\'{\i}}a-Berro}
  et~al.}(2010)\citenamefont{{Garc{\'{\i}}a-Berro}, {Torres}, {Althaus},
  {Renedo}, {Lor{\'e}n-Aguilar}, {C{\'o}rsico}, {Rohrmann}, {Salaris}, and
  {Isern}}}]{garciaberro2010}
\bibinfo{author}{\bibfnamefont{E.}~\bibnamefont{{Garc{\'{\i}}a-Berro}}},
  \bibinfo{author}{\bibfnamefont{S.}~\bibnamefont{{Torres}}},
  \bibinfo{author}{\bibfnamefont{L.~G.} \bibnamefont{{Althaus}}},
  \bibinfo{author}{\bibfnamefont{I.}~\bibnamefont{{Renedo}}},
  \bibinfo{author}{\bibfnamefont{P.}~\bibnamefont{{Lor{\'e}n-Aguilar}}},
  \bibinfo{author}{\bibfnamefont{A.~H.} \bibnamefont{{C{\'o}rsico}}},
  \bibinfo{author}{\bibfnamefont{R.~D.} \bibnamefont{{Rohrmann}}},
  \bibinfo{author}{\bibfnamefont{M.}~\bibnamefont{{Salaris}}},
  \bibnamefont{and} \bibinfo{author}{\bibfnamefont{J.}~\bibnamefont{{Isern}}},
  \bibinfo{journal}{\nat} \textbf{\bibinfo{volume}{465}}, \bibinfo{pages}{194}
  (\bibinfo{year}{2010}), \eprint{1005.2272}.

\bibitem[{\citenamefont{{Winget} et~al.}(2009)\citenamefont{{Winget}, {Kepler},
  {Campos}, {Montgomery}, {Girardi}, {Bergeron}, and {Williams}}}]{winget2009}
\bibinfo{author}{\bibfnamefont{D.~E.} \bibnamefont{{Winget}}},
  \bibinfo{author}{\bibfnamefont{S.~O.} \bibnamefont{{Kepler}}},
  \bibinfo{author}{\bibfnamefont{F.}~\bibnamefont{{Campos}}},
  \bibinfo{author}{\bibfnamefont{M.~H.} \bibnamefont{{Montgomery}}},
  \bibinfo{author}{\bibfnamefont{L.}~\bibnamefont{{Girardi}}},
  \bibinfo{author}{\bibfnamefont{P.}~\bibnamefont{{Bergeron}}},
  \bibnamefont{and}
  \bibinfo{author}{\bibfnamefont{K.}~\bibnamefont{{Williams}}},
  \bibinfo{journal}{Astrophys.\ J.\ Letters} \textbf{\bibinfo{volume}{693}},
  \bibinfo{pages}{L6} (\bibinfo{year}{2009}), \eprint{0901.2950}.

\bibitem[{\citenamefont{{Garc{\'{\i}}a-Berro}
  et~al.}(2011)\citenamefont{{Garc{\'{\i}}a-Berro}, {Torres}, {Renedo},
  {Camacho}, {Althaus}, {C{\'o}rsico}, {Salaris}, and
  {Isern}}}]{garciaberro2011}
\bibinfo{author}{\bibfnamefont{E.}~\bibnamefont{{Garc{\'{\i}}a-Berro}}},
  \bibinfo{author}{\bibfnamefont{S.}~\bibnamefont{{Torres}}},
  \bibinfo{author}{\bibfnamefont{I.}~\bibnamefont{{Renedo}}},
  \bibinfo{author}{\bibfnamefont{J.}~\bibnamefont{{Camacho}}},
  \bibinfo{author}{\bibfnamefont{L.~G.} \bibnamefont{{Althaus}}},
  \bibinfo{author}{\bibfnamefont{A.~H.} \bibnamefont{{C{\'o}rsico}}},
  \bibinfo{author}{\bibfnamefont{M.}~\bibnamefont{{Salaris}}},
  \bibnamefont{and} \bibinfo{author}{\bibfnamefont{J.}~\bibnamefont{{Isern}}},
  \bibinfo{journal}{ArXiv e-prints}  (\bibinfo{year}{2011}),
  \eprint{1107.3016}.

\bibitem[{\citenamefont{{Horowitz} et~al.}(2010)\citenamefont{{Horowitz},
  {Schneider}, and {Berry}}}]{horowitz2010}
\bibinfo{author}{\bibfnamefont{C.~J.} \bibnamefont{{Horowitz}}},
  \bibinfo{author}{\bibfnamefont{A.~S.} \bibnamefont{{Schneider}}},
  \bibnamefont{and} \bibinfo{author}{\bibfnamefont{D.~K.}
  \bibnamefont{{Berry}}}, \bibinfo{journal}{Physical Review Letters}
  \textbf{\bibinfo{volume}{104}}, \bibinfo{pages}{231101}
  (\bibinfo{year}{2010}), \eprint{1005.2441}.

\bibitem[{\citenamefont{{Bildsten} and {Hall}}(2001)}]{bildsten2001}
\bibinfo{author}{\bibfnamefont{L.}~\bibnamefont{{Bildsten}}} \bibnamefont{and}
  \bibinfo{author}{\bibfnamefont{D.~M.} \bibnamefont{{Hall}}},
  \bibinfo{journal}{Astrophys.\ J.\ Letters} \textbf{\bibinfo{volume}{549}},
  \bibinfo{pages}{L219} (\bibinfo{year}{2001}),
  \eprint{arXiv:astro-ph/0101365}.

\bibitem[{\citenamefont{{Deloye} and {Bildsten}}(2002)}]{deloye2002}
\bibinfo{author}{\bibfnamefont{C.~J.} \bibnamefont{{Deloye}}} \bibnamefont{and}
  \bibinfo{author}{\bibfnamefont{L.}~\bibnamefont{{Bildsten}}},
  \bibinfo{journal}{\apj} \textbf{\bibinfo{volume}{580}}, \bibinfo{pages}{1077}
  (\bibinfo{year}{2002}), \eprint{arXiv:astro-ph/0207623}.

\bibitem[{\citenamefont{{Timmes} et~al.}(2003)\citenamefont{{Timmes}, {Brown},
  and {Truran}}}]{timmes2003}
\bibinfo{author}{\bibfnamefont{F.~X.} \bibnamefont{{Timmes}}},
  \bibinfo{author}{\bibfnamefont{E.~F.} \bibnamefont{{Brown}}},
  \bibnamefont{and} \bibinfo{author}{\bibfnamefont{J.~W.}
  \bibnamefont{{Truran}}}, \bibinfo{journal}{Astrophys.\ J.\ Letters}
  \textbf{\bibinfo{volume}{590}}, \bibinfo{pages}{L83} (\bibinfo{year}{2003}),
  \eprint{arXiv:astro-ph/0305114}.

\bibitem[{\citenamefont{{Segretain} and {Chabrier}}(1993)}]{segretain1993}
\bibinfo{author}{\bibfnamefont{L.}~\bibnamefont{{Segretain}}} \bibnamefont{and}
  \bibinfo{author}{\bibfnamefont{G.}~\bibnamefont{{Chabrier}}},
  \bibinfo{journal}{Astronomy\ \&\ Astrophysics}
  \textbf{\bibinfo{volume}{271}}, \bibinfo{pages}{L13+} (\bibinfo{year}{1993}).

\bibitem[{\citenamefont{{Ogata} et~al.}(1993)\citenamefont{{Ogata}, {Iyetomi},
  {Ichimaru}, and {van Horn}}}]{ogata1993}
\bibinfo{author}{\bibfnamefont{S.}~\bibnamefont{{Ogata}}},
  \bibinfo{author}{\bibfnamefont{H.}~\bibnamefont{{Iyetomi}}},
  \bibinfo{author}{\bibfnamefont{S.}~\bibnamefont{{Ichimaru}}},
  \bibnamefont{and} \bibinfo{author}{\bibfnamefont{H.~M.} \bibnamefont{{van
  Horn}}}, \bibinfo{journal}{\pre} \textbf{\bibinfo{volume}{48}},
  \bibinfo{pages}{1344} (\bibinfo{year}{1993}).

\bibitem[{\citenamefont{{Ichimaru} et~al.}(1988)\citenamefont{{Ichimaru},
  {Iyetomi}, and {Ogata}}}]{ichimaru1988}
\bibinfo{author}{\bibfnamefont{S.}~\bibnamefont{{Ichimaru}}},
  \bibinfo{author}{\bibfnamefont{H.}~\bibnamefont{{Iyetomi}}},
  \bibnamefont{and} \bibinfo{author}{\bibfnamefont{S.}~\bibnamefont{{Ogata}}},
  \bibinfo{journal}{Astrophys.\ J.\ Letters} \textbf{\bibinfo{volume}{334}},
  \bibinfo{pages}{L17} (\bibinfo{year}{1988}).

\bibitem[{\citenamefont{{Dewitt} and {Slattery}}(2003)}]{dewitt2003}
\bibinfo{author}{\bibfnamefont{H.}~\bibnamefont{{Dewitt}}} \bibnamefont{and}
  \bibinfo{author}{\bibfnamefont{W.}~\bibnamefont{{Slattery}}},
  \bibinfo{journal}{Contributions\ to\ Plasma\ Physics}
  \textbf{\bibinfo{volume}{43}}, \bibinfo{pages}{279} (\bibinfo{year}{2003}).

\bibitem[{\citenamefont{{Dewitt} et~al.}(1996)\citenamefont{{Dewitt},
  {Slattery}, and {Chabrier}}}]{dewitt1996}
\bibinfo{author}{\bibfnamefont{H.}~\bibnamefont{{Dewitt}}},
  \bibinfo{author}{\bibfnamefont{W.}~\bibnamefont{{Slattery}}},
  \bibnamefont{and}
  \bibinfo{author}{\bibfnamefont{G.}~\bibnamefont{{Chabrier}}},
  \bibinfo{journal}{Physica\ B\ Condensed\ Matter}
  \textbf{\bibinfo{volume}{228}}, \bibinfo{pages}{21} (\bibinfo{year}{1996}).

\bibitem[{\citenamefont{{Segretain}}(1996)}]{segretain1996}
\bibinfo{author}{\bibfnamefont{L.}~\bibnamefont{{Segretain}}},
  \bibinfo{journal}{Astronomy\ \&\ Astrophysics}
  \textbf{\bibinfo{volume}{310}}, \bibinfo{pages}{485} (\bibinfo{year}{1996}),
  \eprint{arXiv:astro-ph/9510118}.

\bibitem[{\citenamefont{{Potekhin}
  et~al.}(2009{\natexlab{a}})\citenamefont{{Potekhin}, {Chabrier}, and
  {Rogers}}}]{potekhin2009}
\bibinfo{author}{\bibfnamefont{A.~Y.} \bibnamefont{{Potekhin}}},
  \bibinfo{author}{\bibfnamefont{G.}~\bibnamefont{{Chabrier}}},
  \bibnamefont{and} \bibinfo{author}{\bibfnamefont{F.~J.}
  \bibnamefont{{Rogers}}}, \bibinfo{journal}{\pre}
  \textbf{\bibinfo{volume}{79}}, \bibinfo{pages}{016411}
  (\bibinfo{year}{2009}{\natexlab{a}}), \eprint{0812.4344}.

\bibitem[{\citenamefont{{Potekhin}
  et~al.}(2009{\natexlab{b}})\citenamefont{{Potekhin}, {Chabrier}, {Chugunov},
  {Dewitt}, and {Rogers}}}]{potekhin2009b}
\bibinfo{author}{\bibfnamefont{A.~Y.} \bibnamefont{{Potekhin}}},
  \bibinfo{author}{\bibfnamefont{G.}~\bibnamefont{{Chabrier}}},
  \bibinfo{author}{\bibfnamefont{A.~I.} \bibnamefont{{Chugunov}}},
  \bibinfo{author}{\bibfnamefont{H.~E.} \bibnamefont{{Dewitt}}},
  \bibnamefont{and} \bibinfo{author}{\bibfnamefont{F.~J.}
  \bibnamefont{{Rogers}}}, \bibinfo{journal}{\pre}
  \textbf{\bibinfo{volume}{80}}, \bibinfo{pages}{047401}
  (\bibinfo{year}{2009}{\natexlab{b}}), \eprint{0909.3990}.

\bibitem[{\citenamefont{{Medin} and {Cumming}}(2010)}]{medin2010}
\bibinfo{author}{\bibfnamefont{Z.}~\bibnamefont{{Medin}}} \bibnamefont{and}
  \bibinfo{author}{\bibfnamefont{A.}~\bibnamefont{{Cumming}}},
  \bibinfo{journal}{\pre} \textbf{\bibinfo{volume}{81}},
  \bibinfo{pages}{036107} (\bibinfo{year}{2010}), \eprint{1002.3327}.

\bibitem[{\citenamefont{Stringfellow et~al.}(1990)\citenamefont{Stringfellow,
  DeWitt, and Slattery}}]{stringfellow1990}
\bibinfo{author}{\bibfnamefont{G.~S.} \bibnamefont{Stringfellow}},
  \bibinfo{author}{\bibfnamefont{H.~E.} \bibnamefont{DeWitt}},
  \bibnamefont{and} \bibinfo{author}{\bibfnamefont{W.~L.}
  \bibnamefont{Slattery}}, \bibinfo{journal}{Phys. Rev. A}
  \textbf{\bibinfo{volume}{41}}, \bibinfo{pages}{1105} (\bibinfo{year}{1990}),
  \urlprefix\url{http://link.aps.org/doi/10.1103/PhysRevA.41.1105}.

\bibitem[{\citenamefont{{Schneider} et~al.}(2012)\citenamefont{{Schneider},
  {Hughto}, {Horowitz}, and {Berry}}}]{schneider2011}
\bibinfo{author}{\bibfnamefont{A.~S.} \bibnamefont{{Schneider}}},
  \bibinfo{author}{\bibfnamefont{J.}~\bibnamefont{{Hughto}}},
  \bibinfo{author}{\bibfnamefont{C.~J.} \bibnamefont{{Horowitz}}},
  \bibnamefont{and} \bibinfo{author}{\bibfnamefont{D.~K.}
  \bibnamefont{{Berry}}}, \bibinfo{journal}{\pre}
  \textbf{\bibinfo{volume}{85}}, \bibinfo{pages}{066405}
  (\bibinfo{year}{2012}).

\bibitem[{\citenamefont{{Horowitz} et~al.}(2007)\citenamefont{{Horowitz},
  {Berry}, and {Brown}}}]{horowitz2007}
\bibinfo{author}{\bibfnamefont{C.~J.} \bibnamefont{{Horowitz}}},
  \bibinfo{author}{\bibfnamefont{D.~K.} \bibnamefont{{Berry}}},
  \bibnamefont{and} \bibinfo{author}{\bibfnamefont{E.~F.}
  \bibnamefont{{Brown}}}, \bibinfo{journal}{\pre}
  \textbf{\bibinfo{volume}{75}}, \bibinfo{pages}{066101}
  (\bibinfo{year}{2007}), \eprint{arXiv:astro-ph/0703062}.

\bibitem[{\citenamefont{{Hughto} et~al.}(2011)\citenamefont{{Hughto},
  {Schneider}, {Horowitz}, and {Berry}}}]{hughto2011}
\bibinfo{author}{\bibfnamefont{J.}~\bibnamefont{{Hughto}}},
  \bibinfo{author}{\bibfnamefont{A.~S.} \bibnamefont{{Schneider}}},
  \bibinfo{author}{\bibfnamefont{C.~J.} \bibnamefont{{Horowitz}}},
  \bibnamefont{and} \bibinfo{author}{\bibfnamefont{D.~K.}
  \bibnamefont{{Berry}}}, \bibinfo{journal}{\pre}
  \textbf{\bibinfo{volume}{84}}, \bibinfo{pages}{016401}
  (\bibinfo{year}{2011}), \eprint{1104.4822}.

\bibitem[{\citenamefont{S.~{Hamaguchi}
  et~al.}(1996)\citenamefont{S.~{Hamaguchi}, {Farouki}, and
  {Dubin}}}]{hamaguchi1996}
\bibinfo{author}{\bibfnamefont{S.}~\bibnamefont{S.~{Hamaguchi}}},
  \bibinfo{author}{\bibfnamefont{R.~T.} \bibnamefont{{Farouki}}},
  \bibnamefont{and} \bibinfo{author}{\bibfnamefont{D.~H.~E.}
  \bibnamefont{{Dubin}}}, \bibinfo{journal}{J. Chem. Phys.}
  \textbf{\bibinfo{volume}{105}} (\bibinfo{year}{1996}).

\bibitem[{\citenamefont{{Steinhardt} et~al.}(1983)\citenamefont{{Steinhardt},
  {Nelson}, and {Ronchetti}}}]{steinhardt83}
\bibinfo{author}{\bibfnamefont{P.~J.} \bibnamefont{{Steinhardt}}},
  \bibinfo{author}{\bibfnamefont{D.~R.} \bibnamefont{{Nelson}}},
  \bibnamefont{and}
  \bibinfo{author}{\bibfnamefont{M.}~\bibnamefont{{Ronchetti}}},
  \bibinfo{journal}{\prb} \textbf{\bibinfo{volume}{28}}, \bibinfo{pages}{784}
  (\bibinfo{year}{1983}).

\bibitem[{\citenamefont{{ten Wolde} et~al.}(1996)\citenamefont{{ten Wolde},
  {Ruiz-Montero}, and {Frenkel}}}]{tenwolde96}
\bibinfo{author}{\bibfnamefont{P.~R.} \bibnamefont{{ten Wolde}}},
  \bibinfo{author}{\bibfnamefont{M.~J.} \bibnamefont{{Ruiz-Montero}}},
  \bibnamefont{and}
  \bibinfo{author}{\bibfnamefont{D.}~\bibnamefont{{Frenkel}}},
  \bibinfo{journal}{\jcp} \textbf{\bibinfo{volume}{104}}, \bibinfo{pages}{9932}
  (\bibinfo{year}{1996}).

\bibitem[{\citenamefont{{Althaus} et~al.}(2012)\citenamefont{{Althaus},
  {Garc{\'{\i}}a-Berro}, {Isern}, {C{\'o}rsico}, and {Miller
  Bertolami}}}]{althaus2012}
\bibinfo{author}{\bibfnamefont{L.~G.} \bibnamefont{{Althaus}}},
  \bibinfo{author}{\bibfnamefont{E.}~\bibnamefont{{Garc{\'{\i}}a-Berro}}},
  \bibinfo{author}{\bibfnamefont{J.}~\bibnamefont{{Isern}}},
  \bibinfo{author}{\bibfnamefont{A.~H.} \bibnamefont{{C{\'o}rsico}}},
  \bibnamefont{and} \bibinfo{author}{\bibfnamefont{M.~M.} \bibnamefont{{Miller
  Bertolami}}}, \bibinfo{journal}{Astronomy\ \&\ Astrophysics}
  \textbf{\bibinfo{volume}{537}}, \bibinfo{eid}{A33} (\bibinfo{year}{2012}),
  \eprint{1110.5665}.

\end{thebibliography}

\end{document}